\documentclass[10pt,letterpaper]{article}
\usepackage[top=2.75cm,left=2.75cm,footskip=1cm,marginparwidth=1.5cm]{geometry}
\usepackage[table]{xcolor}
\usepackage{array}
\newcolumntype{H}{>{\setbox0=\hbox\bgroup}c<{\egroup}@{}}

\usepackage{hyperref}
\usepackage[utf8]{inputenc}
\usepackage{graphicx}
\usepackage{amsmath}
\usepackage{amssymb}
\usepackage{booktabs} %toprule e bottonrule
\usepackage{multirow}
\usepackage{setspace} %configurar espaçamento entre linhas
\usepackage{pgfplots}
\usepackage{framed}

\usepackage[ruled, boxed, linesnumbered, vlined, english]{algorithm2e}
\usepackage{booktabs}
\usepackage{multirow}
\usepackage{epstopdf}
\usepackage[flushleft]{threeparttable} %notas na tabela
\usepackage{titlesec}\titleformat{\subsubsection}[runin]{\bf}{}{}{}[]

\usepackage{csvsimple}
\usepackage{siunitx}
  \title{Branch-and-cut and iterated local search for the weighted $k$-traveling repairman problem: an application to the maintenance of speed cameras}
  
\begin{document}

    \begin{flushleft}
    	{\Large
    		\textbf\newline{{Branch-and-cut and iterated local search for the weighted $k$-traveling repairman problem: an application to the maintenance of speed cameras}}
    	}
    	\newline
    	% authors go here:

    		Albert Einstein Fernandes Muritiba \textsuperscript{1},	\texttt{einstein@ufc.br}\\
    		Tib\'erius O. Bonates\textsuperscript{1}, \texttt{tb@ufc.br}\\
    		Stênio Oliveira Da Silva \textsuperscript{2}, \texttt{stenio.wow@gmail.com}\\
    		Manuel Iori \textsuperscript{3}, \texttt{manuel.iori@unimore.it}\\
    	\bigskip

    	\bf{1} Department of Statistics and Applied Mathematics, Federal University of Ceará, Brazil \\
    	\bf{2} Mestrado Integrado Profissional em Computação Aplicada,	State University of Ceará, Brazil    	\\
  		\bf{3} Department of Sciences and methods for Engineering, 	University of Modena and Reggio Emilia, Italy\\
   	
    	\bigskip
    	* correseponding author: einstein@ufc.br
    \end{flushleft}

\section*{Abstract}
    	Private enterprises and governments around the world use speed cameras to control traffic flow and limit speed excess. Cameras may be exposed to difficult weather conditions and typically require frequent maintenance. When deciding the order in which maintenance should be performed, one has to consider both the traveling times between the cameras and the traffic flow that each camera is supposed to monitor. In this paper, we study the problem of routing a set of technicians to repair cameras by minimizing the total weighted latency, that is, the sum of the weighted waiting times of each camera, where the weight is a parameter proportional to the monitored traffic. The resulting problem, called weighted k-traveling repairman problem (wkTRP), is a generalization of the well-known traveling repairman problem and can be used to model a variety of  real-world applications. To solve the wkTRP, we propose an iterated local search heuristic and an exact branch-and-cut algorithm enriched with valid inequalities. The effectiveness of the two methods is proved by extensive computational experiments performed both on instances derived from a real-world case study, as well as on benchmark instances from the literature on the wkTRP and on related problems.
    %
    
    % Sample
    %\KEYWORDS{deterministic inventory theory; infinite linear programming duality; 
    %  existence of optimal policies; semi-Markov decision process; cyclic schedule}
    
    % Fill in data. If unknown, outcomment the field
%    \keywords{Traveling repairman problem \and Weighted latency \and Speed cameras \and Branch-and-cut}

%    \maketitle

%%%%%%%%%%%%%%%%%%%%%%%%%%%%%%%%%%%%%%%%%%%%%%%%%%%%%%%%%%%%%
\section{Introduction}
\begin{figure}[b]
   \centering
   \includegraphics[width=0.75\linewidth]{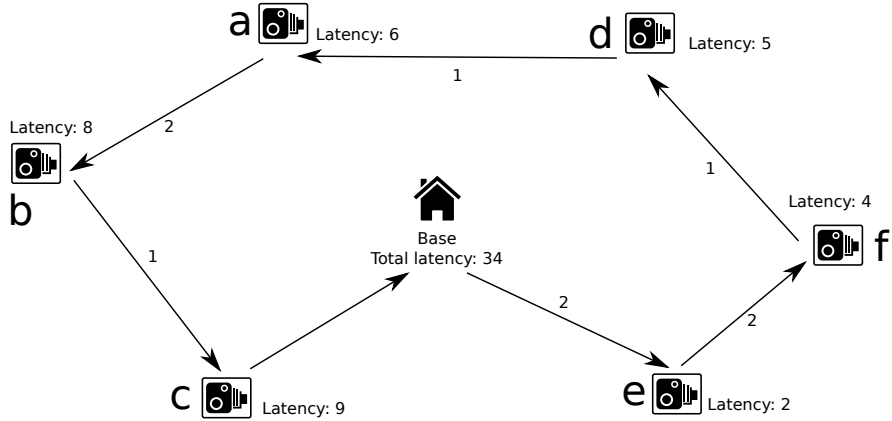}	
   \caption{An example of the traveling repairman problem.}
   \label{fig:pml}
\end{figure}

The {\em traveling repairman problem} (TRP) requires the determination of the visiting sequence of a set of clients by means of a single repairman, such that the sum of latencies (i.e., waiting times) of all clients is minimized. The TRP has been extensively studied in the literature because it can be used to model a variety of real-world situations and also because it is a difficult problem to solve in practice. As a consequence of the large number of studies, the TRP can be found under different names, such as the minimum latency problem, the deliveryman problem, and the traveling salesman problem with cumulative costs, among others (see, e.g., \cite{Minieka1989, Bianco1993, Sitters2002, Ngueveu2010}). We adopt the TRP nomenclature because we believe it fits more precisely the real-world application that we consider in this work.
Figure \ref{fig:pml} shows a TRP example involving a depot and six clients, where the repairman visits the clients according to the sequence (e--f--d--a--b--c) and obtains a cumulative latency of 2+4+5+6+8+9=34 time units. 

The TRP was introduced in \cite{conway1967} as a variant of the {\em machine scheduling problem}, but has since been frequently described as a variant of the {\em traveling salesman problem} (TSP) (see, e.g., \cite{Picard1978}). While the TRP may appear similar to the TSP, some important features distinguish the two problems: (i) in the TSP, cost is related to the salesman trajectory, whereas in the TRP the cost concerns the times in which clients are serviced; (ii) a  local change in the visiting sequence has limited effect on a TSP circuit, as compared to what happens in a TRP solution, in which even small changes can have a global effect on the overall cost; (iii) the TSP is trivial when clients are located on a tree, whereas the TRP remains difficult (\cite{Minieka1989}); (iv) a TSP solution cost is independent of the starting client, whereas the TRP either has a predefined starting point, or the definition of the starting point is an optimization problem in itself (see, e.g., \cite{Wu2000}).

\begin{figure}[tbph]
    \centering
    \includegraphics[width=0.9\linewidth]{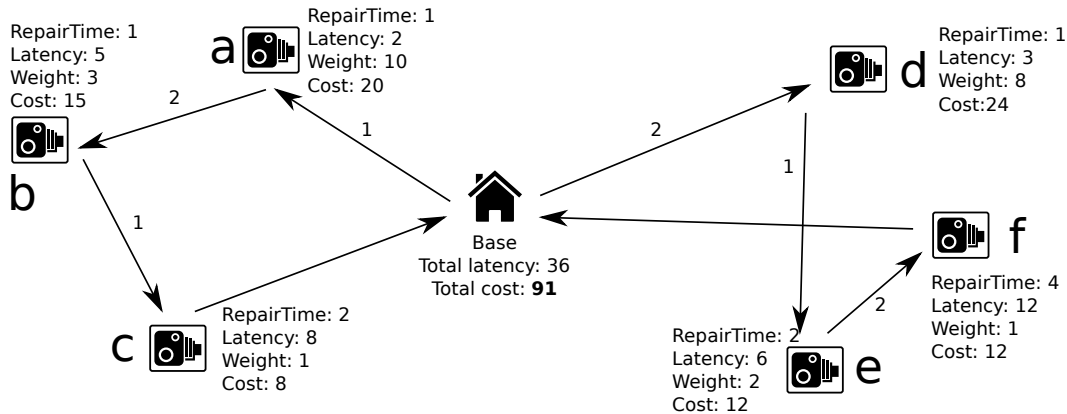}
    \caption{An example of the {\em weighted $k$-traveling repairman problem} (wkTRP).}
    \label{fig:kpmlpr}
  \end{figure}

In this paper, we study the {\em weighted $k$-traveling repairman problem} (wkTRP), a generalization of the TRP in which: (i) $k$ repairmen can be used to visit the clients; (ii) service times are associated with each client; (iii) an input weight is assigned to each client; and (iv) the objective is to minimize the total weighted latency. The latency of a client is the time elapsed between the moment in which the repairmen started his route, added to the client's repair time.
Figure \ref{fig:kpmlpr} shows an example of a wkTRP solution with two repairmen, six clients, and one starting point (named ``Base'' in the figure). The cost of each client is calculated as the product between sher latency and her weight. The total weighted latency is thus 12+12+24+20+15+8=91 (whereas the total latency is 36).     

The wkTRP can be used to model delivery problems, as well as maintenance routing problems, where a certain information on the importance of each client is provided as a weight, and weighted waiting times are to be minimized by using one or more technicians.
Apart from these two classical applications, the wkTRP can also de used to model a number of different real-world applications, including: disk-head schedule, elevator dispatching, pharmaceutical distribution, and autonomous Web crawlers optimization (see, e.g., \cite{Blum1994, Ausiello2000, Fakcharoenphol2003}). 

Our interest in the wkTRP is motivated by a real-world case study for repairing speed cameras in the area of Rio de Janeiro, Brazil. The problem can be modeled as a wkTRP instance in which each service time corresponds to the time required to repair a broken camera, and each weight is proportional to the amount of traffic that is expected to be monitored by each broken camera. In order to solve the wkTRP, we propose an iterated local search heuristic, which works quite well on a set of real-world instances, and a branch-and-cut method enriched with several valid inequalities, which solves to optimality a large number of benchmark instances of the wkTRP and of similar problems from the literature.

The remainder of the paper is organized as follows. In Section \ref{sec:literature}, we provide a quick overview of the related literature. Section \ref{sec:case} describes the details of the real-world application. In Section \ref{sec:wkTRP}, we introduce a wkTRP formulation, along with three families of strengthening cutting planes. Section \ref{sec:heuristic} describes our proposed iterated local search algorithm, whereas Section \ref{sec:results} is devoted to discussing the outcome of extensive computational tests. Indeed, since the wkTRP is quite general, we do not only solve wkTRP instances, but also address a number of problem variants from the literature. We provide our conclusions in Section \ref{sec:conclusions}, where we also outline directions for further research.

%%%%%%%%%%%%%%%%%%%%%%%%%%%%%%%%%%%%%%%%%%%%%%%%%%%%%%%%%%%%%
\section{Problem Description and Literature Review}\label{sec:literature}
Assume that we are given a set of $n$ clients, a single depot, and $K$ repairmen. We consider a complete directed graph $G=(V,A)$, where $V = \{0, 1, \dots, n\}$ comprises the depot (identified with vertex 0) and the clients (vertices from 1 to $n$). Arcs $(i,j) \in A$ are associated with a non-negative traveling time $c_{ij}$. Each client $i$ has a service time $r_i$ and a priority weight $w_i$. The wkTRP requires the definition of a set of $K$ routes (one for each repairman) that start and end at the depot, such that each client is visited by exactly one repairman.
We assume that all repairmen leave the depot simultaneously. We define the latency (i.e., the waiting time) of a client as the time elapsed between the moment when the repairmen leaves the depot and the moment when the repairman leaves the client's location (in our real-world application, the latency period ends when the camera has been repaired). The service time of a client is, therefore, part of the client's latency. We note that some authors, instead, do not regard the service time of a client as part of her latency (see, e.g., \cite{Tsitsiklis1992,Jothi2007}). The two variants can be easily interchanged by means of a matrix transformation in which the service time $r_i$ is removed from vertex $i$ and added either to the costs of the incoming arcs $(h,i) \in A$ or to the costs of the outgoing arcs $(i,j) \in A$. Finally, the wkTRP objective is to minimize the total weighted latency over all clients.

The wkTRP generalizes the {\em $k$-traveling repairman problem} (kTRP) -- in which all clients have equal weights --, the {\em weighted traveling repairman problem} (wTRP) -- in which a single repairman is available --, as well as the previously mentioned TRP. We are not aware of any work addressing the wkTRP, but the literature on the TRP, and on many TRP variants, is vast. Providing a complete survey of this literature is out of the scope of this paper; nevertheless, we present some pointers to the most recent and relevant publications. 

The TRP is known to be NP-hard for general distance matrices (\cite{Sahni1976}). \cite{Wu2000} showed that it can be solved in polynomial time on some special graphs via the use of dynamic programming. The author also presented algorithms to find the optimal departure vertex, both for the TRP and for the kTRP. Despite admitting some polynomially solvable classes, the TRP remains NP-hard even in the case of weighted trees, as shown in \cite{Sitters2002}.

A number of polynomial-time approximation schemes have been devised in the TRP literature. To the best of our knowledge, the most recent ones are due to \cite{Archer2008}, who obtained a TRP approximation algorithm by using a limited number of calls to an inner approximation scheme for the prize-collecting Steiner tree, and to \cite{Nagarajan2008}, who focused on the case of asymmetric distances. \cite{Fakcharoenphol2007} dealt, instead, with the kTRP and proposed an approximation algorithm that requires finding the least-cost rooted tree spanning $i$ vertices.

A large number of mathematical formulations and exact algorithms have been proposed to solve the TRP. A good survey and classification of early works was provided by \cite{GV95}, who also described dominance and equivalence relationships among the formulations. Later, \cite{Wu2004} proposed an exact algorithm based on dynamic programming, branch-and-bound, and non-trivial lower bounding functions, consistently solving instances with up to 25 clients.
\cite{mendez2008} introduced a formulation in which a binary variable $x_{ij}$ indicates whether arc $(i,j)$ is used, and a continuous variable $y_{ij}^k$ indicates whether arc $(i,j)$ is used in the path from 0 to $k$. They presented results for instances involving up to 40 vertices. \cite{Angel-Bello2013} presented two TRP formulations using a variable $x_i^k$ indicating if client $i$ is in position $k$ in the route, and a variable $y_{ij}^k$ indicating whether or not client $i$ is in position $k$ and client $j$ is in position $k+1$. They also solved instances with up to 25 clients.
\cite{AFPU1395} studied the polytope associated with the formulation by \cite{Picard1978}, which is also based on the $y_{ij}^k$ variable just mentioned. They solved instances with up to 100 vertices using a dedicated branch-cut-and-price algorithm. The formulation by \cite{Picard1978} was also improved by \cite{MMZ14}, this time with the use of a branch-and-cut algorithm. \cite{RM14} modeled the TRP as a single-route set partitioning problem and then solved it by using a dynamic ng-path relaxation, obtaining some improvements with respect to the results of \cite{AFPU1395}. In a recent technical report, \cite{BSU17} obtained further improvements with the use of a branch-and-price algorithm that relies on a number of techniques, including a labeling algorithm with an enhanced dominance rule, generalized ng-paths, and reduced cost fixing.

As far as exact methods for the kTRP are concerned, \cite{OKD17} generalized a TRP formulation by \cite{Angel-Bello2013} to the kTRP case and tested it on instances involving up to 50 nodes and 10 vehicles. \cite{Luo2014} 
developed a branch-and-price-and-cut for the kTRP variant in which each route is constrained by a maximum traveling distance. The authors used a label-setting algorithm with bidirectional search to solve the pricing subproblems.

In terms of metaheuristic methods for the TRP, we mention the combination of {\em greedy randomized adaptive search procedure} (GRASP), {\em variable neighborhood descent} (VND), and {\em variable neighborhood search} (VNS) by \cite{Salehipour2011}, the combination of GRASP, iterated local search and randomized VND by \cite{MSVO12}, and the VNS algorithm by \cite{Mladenovic2013}.

In the standard three-field notation for scheduling problems, the TRP can be denoted as $1|s_{ij}|\sum C_j$ (one machine, sequence dependent setup times, minimization of total completion time), while the wkTRP can be denoted as $P|s_{ij}|\sum w_j C_j$ (parallel identical machines, sequence dependent setup times, minimization of total weighted completion time).
Some papers made use of this scheduling representation of the problem to devise mathematical models. That was the case, for example, of \cite{BGS08}, who studied single machine scheduling problems with sequence-dependent setup times, solving the TRP and the $1|s_{ij}|\sum T_j$ (total tardiness minimization) through the use of mathematical programming formulations and an exact branch-and-bound algorithm. Differently, \cite{TA13} solved the $1|s_{ij}|\sum w_j T_j$ by means of an exact method based on Lagrangian relaxation and dynamic programming.

Other interesting variants of the TRP have been addressed in the literature. \cite{Tsitsiklis1992} studied the TRP with time windows and service times and obtained several complexity results. \cite{BVP13} presented a VNS for the kTRP with heterogeneous fleet and time windows. \cite{Ngueveu2010} developed a memetic algorithm for the cumulative capacitated vehicle routing problem, that is, the variant of the kTRP in which each client has a weighted demand and each vehicle has a maximum capacity. The same problem was later addressed by \cite{Ribeiro2012} via an adaptive large neighborhood search, and by \cite{NMAC18} with mathematical models and an iterated greedy procedure.

Dynamic problems in which information is revealed during optimization have also been addressed. The dynamic TRP was solved, among others, by \cite{KPPS03} and \cite{JW06}. Another dynamic problem originating from a real-world application was studied by \cite{LG13}. They considered the signal repair problem, which includes inspection and repair of traffic control devices such as signal lights, traffic controllers, and vehicle detectors, in the city of Kaohsiung (China). They modeled the problem as a kTRP and solved it with an online simulated annealing algorithm. Their problem is the one that shares more similarities with our real-world application, but it is still distinct from our case due to the fact that it does not take into account client weights.

Another related problem having a precise real-world application is the {\em workover rig routing problem} (WRRP). The WRRP arises in the operations of onshore oil fields, where a set of workover rigs located at different positions must service oil wells requesting maintenance as soon as possible. It is important to service the wells in a timely fashion to minimize production loss, so the WRRP objective is equal to the sum of arrival times at the wells multiplied by the production loss rates. The WRRP objective is thus equivalent to that of the wkTRP, although the WRPP additionally considers multiple depots and time windows. Efficient metaheuristics for the WRRP have been proposed by \cite{RLM12}, whereas an updated survey can be found in \cite{BCR16}. A related problem focused at the minimization of risk during helicopter transportation was addressed by \cite{QGH11}.

\section{Case Study}\label{sec:case}

\begin{figure}[ht]
    \centering
    \includegraphics[width=0.7\linewidth]{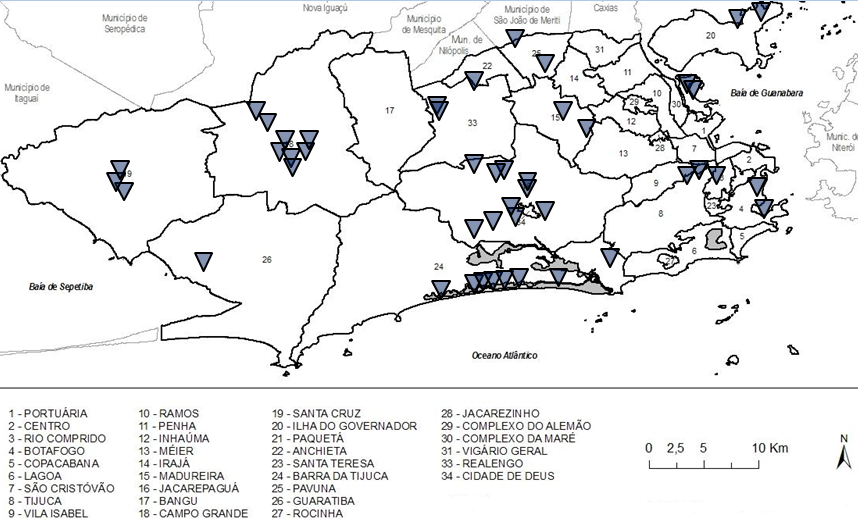}
    \caption{Speed cameras in the city of Rio de Janeiro.}
    \label{fig:MAP_RJ}
\end{figure}

This work is motivated by a practical application arising from the operations of a Brazilian company -- which shall remain anonymous in this paper, as per their request -- whose main business activities consist of the assembly, deployment, and maintenance of speed cameras.
Our case study is focused on the city of Rio de Janeiro, where the company is responsible for the maintenance of 50 speed cameras. The scenario is represented in Figure \ref{fig:MAP_RJ}, where speed cameras are marked with triangles.

Speed cameras are geographically-dispersed devices that play an important role in traffic safety by inhibiting dangerous behaviours from drivers. These devices can detect and record a number of transit infractions, which are turned into fines.
An interesting aspect about speed cameras is their maintenance. Speed cameras demand frequent maintenance because they can be exposed to harsh weather conditions (sun, rain, etc). The diagnosis of certain malfunctions can be done by means of a remote communication system, but reparations must be performed on the spot. 
A base team and up to three repairman teams are responsible for the maintenance of the cameras in the metropolitan area of Rio de Janeiro. While the base team works at the office, the repairman teams travel to reach the cameras and perform the required reparations. 

The contract in force between the company and the city transit authority states that a fraction of the fines that are issued by the speed cameras is earned by the company itself. In this way, a malfunctioning device implies a monetary loss for the company. 
An additional cost is directly imposed on the company by the transit authority in case a camera is not working. Thus, the resulting overall cost of a malfunctioning speed camera is proportional to its inoperative time and to the frequency with which the camera issues fines. Therefore, the main objective of the maintenance plan is the minimization of the total cost induced by the inoperative time of the defective devices.
        	
This case study can be directly modeled as a wkTRP instance by considering the inoperative cameras as the clients to be visited. The value of $K$ is set to be the number of available repairman teams. The device remote diagnosis allows for the estimation of the repair time (service time) needed to perform each reparation activity. The client weights are calculated on the basis of the historic data of each speed camera. %, and are set to be equal to the average cost of the fines historically issued per minute by each camera.
The total weighted latency, therefore, corresponds to the expected cost caused by the inoperability of the cameras.

\section{Mathematical Model and Branch-and-Cut}\label{sec:wkTRP}
In this section, we propose a novel model for the wkTRP, referred to as wkTRPM, which is based on the concept of \emph{multi-commodity flows}. The model, presented in Section \ref{subsec:model}, originates from the one proposed for the TRP by \cite{mendez2008}, but is extended to take into account  $K$ routes and the weighted latency minimization. Furthermore, it is enriched by a number of valid inequalities, either newly proposed or adapted from existing inequalities in the literature, which are added on the fly in a branch-and-cut fashion. We describe the valid inequalities in Section \ref{subsec:valid-inequalities}, their separation in Section \ref{subsec:separation}, and the branch-and-cut framework in Section \ref{subsec:branch-and-cut}.

\subsection{Mathematical Model}\label{subsec:model}

The mathematical model that we introduce makes use of three families of binary variables:
\begin{itemize}
\item variable $z_{ij}$ takes the value 1 if client $i$ is visited before (even if not immediately) client $j$ in a route, and takes value 0 otherwise, for $i,j = 1, \ldots n, i\neq j$;
\item variable $f^k_{ij}$ takes the value 1 if arc $(i,j)$ is used in the path between the depot and client $k$, and takes value 0 otherwise, for $i,j,k = 0, \ldots n, ~j\neq i\neq k, ~j, k > 0$;
\item variable $s_{ij}$ takes the value 1 if clients $i$ and $j$ are serviced in the same route, and takes value 0 otherwise, for $i,j = 1, \ldots n, i<j$.
\end{itemize}
Variables $z_{ij}$ are adopted to model the vehicle flow, imposing precedences among clients visited in the same routes, whereas $f^k_{ij}$ are the multi-commodity variables that prevent the creation of subtours. Variables $s_{ij}$ are auxiliary variables that are used within valid inequalities. 

The wkTRP can then be modeled as follows.
{ 
\begin{equation}
\text{(wkTRPM) \qquad minimize }   \sum_{i=0}^{n}\sum_{j=1, i\neq j}^{n}\left(c_{ij}+r_j\right) \sum_{k=1, k \neq i}^{n} w_k f^k_{ij}  \label{eq:mkpmlr:fo} 
\end{equation}
subject to
\allowdisplaybreaks
\begin{align}
  	 & z_{ij} + z_{ji} = s_{ij},                         & \forall i,j = 1,\ldots,n, \; i<j \label{eq:mkpmlr:c1}                 \\
  	 & z_{jk} = \sum_{i=0, i \notin \{j,k\}}^{n}f^k_{ij}, & \forall j,k = 1,\ldots,n, ~j\neq k \label{eq:mkpmlr:c4}               \\
  	 & z_{jk} = \sum_{i=1, i \neq j}^{n}f^k_{ji},        & \forall j,k = 1,\ldots,n, ~j\neq k \label{eq:mkpmlr:c5}               \\
  	 & {\sum_{i=1}^{n}} f^j_{0i} = 1,             & \forall j = 1,\ldots,n \label{eq:mkpmlr:c5.1}                         \\
  	 & \sum_{j=1}^{n}f^j_{0j} = K,                        & \label{eq:mkpmlr:c6}                                                  \\
  	 & \sum_{j=1}^{n}f^0_{j0} = K,                        & \label{eq:mkpmlr:c6.1}                                                \\
  	 & \sum\limits_{i=0, i\neq j}^{n} f^j_{ij} = 1,      & \forall j = 1,\ldots,n \label{eq:mkpmlr:c7}                           \\
  	 & {\sum\limits_{j=0, i\neq j}^{n}} f^j_{ij} = 1,      & \forall i = 1,\ldots,n \label{eq:mkpmlr:c8}                           \\
  	 &{\sum_{k=1,i \notin \{j,k\}}^{n}f^k_{ij}}  \leq  (n-K)f^j_{ij},      & \forall i,j = 0, \ldots n, ~j\neq i, ~j > 0 \label{eq:mkpmlr:c9S}    \\
  	 & s_{ij} \in \{0,1\},                               & \forall i,j = 1,\ldots,n, ~i < j \label{eq:mkpmlr:bin1}                    \\
  	 & z_{ij} \in \{0,1\},                               & \forall i,j = 1,\ldots,n, ~i \neq j \label{eq:mkpmlr:bin2}                       \\
  	 & f^k_{ij} \in \{0,1\},                             & \forall i,j,k = 0, \ldots n, ~j\neq i\neq k, ~j, k > 0 \label{eq:mkpmlr:bin3}
\end{align}
}
The objective function \eqref{eq:mkpmlr:fo} requires the minimization of the total weighted latency. Constraints \eqref{eq:mkpmlr:c1} impose $s_{ij} = 1$ if and only if clients $i$ and $j$ are serviced by the same repairman. At the same time, these constraints prevent subtours of cardinality 2. Constraints \eqref{eq:mkpmlr:c4} ensure that client $j$ is visited before client $k$ (i.e., $z_{jk}=1$) if and only if one of the incoming arcs of $j$ is used in the path to $k$.
Constraints \eqref{eq:mkpmlr:c5} impose a similar relation between $z_{jk}$ and, this time, the arcs that leave $j$. Constraints \eqref{eq:mkpmlr:c5.1} impose that the depot precedes any other vertex in the graph, whereas constraints \eqref{eq:mkpmlr:c6} and \eqref{eq:mkpmlr:c6.1} ensure, respectively, that exactly $K$ vehicles leave and return to the depot. Constraints \eqref{eq:mkpmlr:c7} and \eqref{eq:mkpmlr:c8} enforce, respectively, that each client has one {incoming} and one {outgoing} arc.
Constraints \eqref{eq:mkpmlr:c9S} allow the use of arc $(i,j)$ to reach any other successive client $k$ only if the arc has actually been selected (i.e., if $f^j_{ij}=1$). The domains of the decision variables are defined by constraints \eqref{eq:mkpmlr:bin1}--\eqref{eq:mkpmlr:bin3}. 

We bring attention to to some facts about the wkTRPM model given by \eqref{eq:mkpmlr:fo}--\eqref{eq:mkpmlr:bin3}:
\begin{itemize}
    \item The model has polynomial size, with $\mathcal{O}(n^3)$ variables and $\mathcal{O}(n^2)$ constraints;
    \item As previously mentioned, $s_{ij}$ are auxiliary variables to be used within valid inequalities, so replacing \eqref{eq:mkpmlr:c1} and \eqref{eq:mkpmlr:bin1} by $z_{ij} + z_{ji} \le 1$ would still lead to a correct wKTRP formulation;
\item Classical multi-commodity flow formulations are usually based on continuous flow variables. In our proposed model, we can replace \eqref{eq:mkpmlr:bin3} by $f^k_{ij} \geq 0$ and still obtain a valid wkTRP model. This is due to the presence of \eqref{eq:mkpmlr:c4} and \eqref{eq:mkpmlr:c5}. We opted for maintaining the binary variable requirement because this led to better computational results, probably due to a deeper use of preprocessing techniques by the commercial solver invoked to solve the model;
\item Subtours are prevented by the combined effect of \eqref{eq:mkpmlr:c4}, \eqref{eq:mkpmlr:c5} and \eqref{eq:mkpmlr:c6}, which impose the existence of a path from the depot to all clients, and by \eqref{eq:mkpmlr:c7}, \eqref{eq:mkpmlr:c8} and \eqref{eq:mkpmlr:c9S}, which prohibit subtours that do not pass through the depot by requiring exactly one incoming arc to each client.
\end{itemize}
        
\subsection{Valid Inequalities}\label{subsec:valid-inequalities}
    
We strengthen model wkTRPM by using the following three families of valid inequalities.

\subsubsection*{$f$-activation cuts.}~ We recall that $f^j_{ij}=1$ implies arc $(i,j)$ is used, therefore $f^j_{ij}=1$ if $(i,j)$ is in the path to any subsequent client $k$, and $f^j_{ij}=0$ otherwise. This is guaranteed by constraints \eqref{eq:mkpmlr:c9S}. A valid inequality, originally presented by \cite{mendez2008} for the TRP, can be obtained by disaggregating the summation in the left-hand side of \eqref{eq:mkpmlr:c9S}, thus obtaining:
\begin{align}
 & f^k_{ij}  \leq  f^j_{ij}                          & \forall i,j,k = 0, \ldots n, ~i\neq j, j\neq k, i\neq k, ~j,k > 0 \label{eq:mkpmlr:c9}     
\end{align}

\subsubsection*{$z$-activation cuts.}~ In the cases in which either arc $(i,j)$ is in the path to client $k$, or arc $(k,i)$ is  in the path to $j$, or arc $(i,k)$ is in the path to $j$, then $i$ precedes $j$. In addition, these three conditions are mutually exclusive (provided there are no subtours). We can thus state:
 \begin{align}
 & f^k_{ij} +f^j_{ki} +f^j_{ik} \le z_{ij}   & \forall i,j,k = 1,\ldots,n,\; ~i\neq j\neq k, i\neq k \label{eq:mkpmlr:c14}           \end{align}

\subsubsection*{Pigeonhole cuts.}~
Let us consider a wkTRP with $K$=4 repairmen. If we select any subset of five clients, then there surely exists a pair of those clients that is serviced by the same repairman. Therefore, the sum of all $s_{ij}$ variables for $i$ and $j$ in this subset is at least one. This follows directly from the pigeonhole or Dirichlet's box principle, and can be used to provide a valid inequality for the wkTRP as follows.

Let $\bar G=(\bar V, \bar E)$ be the undirected support graph induced by an integer feasible solution $\bar s$ to model wkTRPM, with $\bar V \subset \{1, \ldots, n\}$ and $\bar E=\{(i,j) \in \bar V^2: \bar s_{ij}=1\}$. If $|\bar V| > K$, then $\bar G$ is composed by at least one and at most $K$ disjoint cliques. Such cliques result from the transitivity property of the $s$-variables. 
Function $\omega$, given in Equation \eqref{eq:omega}, estimates the minimum number of edges, and thus of $s_{ij}$ variables taking the value 1, in $\bar G$. The function takes as argument the cardinality of $\bar V$, expressed as $\gamma = |\bar V|$, and the number $K$ of repairmen, and returns the number of edges as:
 {
     \begin{equation}
    \omega(\gamma,K) = (\gamma\mod{K})\cdot \tau\left(\left\lceil \frac{\gamma}{K} \right\rceil\right) +
                 (K - (\gamma\mod{K}) )\cdot\tau\left(\left\lfloor \frac{\gamma}{K} \right\rfloor\right)  \label{eq:omega}    
     \end{equation}     
     where  $\tau(x) = \frac{x^2-x}{2} $ is the number of edges in a clique with $x$ vertices.
         }
         
\begin{figure}[ht]
                   \centering
                   \includegraphics[width=0.7\linewidth]{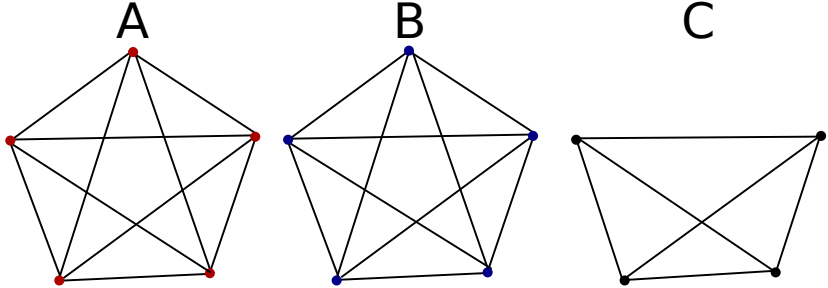}
                   \caption{Graph partition -- $\omega$ function.}
                   \label{fig:omega}
                \end{figure}

{Figure \ref{fig:omega} shows an example of a graph $\bar G$ having $\gamma=14$ vertices distributed in $K=3$ disjoint cliques (A, B, and C). There are $14\mod{3} = 2$ cliques having $\left\lceil \frac{14}{3} \right\rceil = 5$ vertices and $3 - (14\mod{3}) = 1$ clique having $\left\lfloor \frac{14}{3} \right\rfloor = 4$ vertices.} This partitioning pattern where $\bar G$ contains $K$ cliques with cardinality between $\left\lfloor \frac{\gamma}{K} \right\rfloor$ and $\left\lceil \frac{\gamma}{K} \right\rceil$ is the one with the minimum number of edges, namely $\omega(14,3) = 2\times\tau(5) + 1\times\tau(4) = 2\times10 + 1 \times 6 = 26$ edges in this example.
Indeed, any partitioning of $\bar G$ that contains a clique with cardinality greater than $\left\lceil \frac{\gamma}{K} \right\rceil$ does not lead to the minimum possible number of edges, because there would be another $\bar G$ partitioning in which a vertex belonging to the largest clique is moved to a smaller clique, thus resulting in a decrease in the number of edges in $\bar G$. Therefore, there is always a clique smaller than $\left\lceil \frac{\gamma}{K} \right\rceil$ or $\bar G$ has less than $K$ cliques, and in this case a vertex can be removed from one of the cliques to unlock a new clique, thus reducing the number of edges.
        
Function $\omega$ can therefore be used to bound by below the minimum number of $s$-variables taking the value 1 in a given solution, as follows:
        \begin{align}
         & \sum_{i,j \in \bar V; i<j} s_{ij} \geq \omega(|\bar V|,K)               & \forall~ \bar V \subset \{1,\ldots,n\}, |\bar V| > K. \label{eq:mkpmlr:c10}
         \end{align}

\subsection{Separation procedures}\label{subsec:separation}

In our framework, we separate the violated inequalities \eqref{eq:mkpmlr:c9} and \eqref{eq:mkpmlr:c14} through complete enumeration, as this can be done in polynomial time. The set of inequalities \eqref{eq:mkpmlr:c10}, instead, cannot be efficiently separated by enumeration since its cardinality is $\mathcal{O}(2^n)$.
The separation problem for \eqref{eq:mkpmlr:c10} is to find, for any $\gamma \in \{(K+1), \dots, N\}$, a subset $\bar V \subset \{1, \dots, n\}$ having cardinality $\gamma$ and minimizing $\zeta = \sum_{i,j\in \bar V:i < j} s_{ij}$. If $\zeta < \omega(\gamma,K)$, then $\bar V$ corresponds to a violation of \eqref{eq:mkpmlr:c10}. 

This problem can be reduced to the \emph{maximum diversity problem}, that is known to be NP-hard (see \cite{Kuo1993,Ghosh1996}).
In view of such computational difficulty, we propose the greedy heuristic separation procedure that is outlined in Algorithm \ref{alg:pigeon}. For each client $i \in \{1, \dots, n\}$, we initialize the set $\bar V$ to $\{i\}$ and a parameter $\delta$ to 0. Then, we repeatedly add to $\bar V$ a client $j^* \notin \bar V$ whose inclusion in $\bar V$ minimizes the resulting value of $\zeta$, i.e., $j^* \in \mbox{argmin}_{k \notin \bar V}\left\{\sum_{i,j\in \bar V\cup\{k\}; i < j} \bar s_{ij}\right\}$. 
Whenever $\zeta+ \delta +\epsilon < \omega(|\bar V|,K)$ holds, then we have found a violated inequality of type \eqref{eq:mkpmlr:c10}. In this case, we add said inequality to the model, set $\delta =\omega(|\bar V|,K) - \zeta$ and repeat the process. The amount of violation of a separated cut is stored in $\delta$ at each iteration. This value limits the separation of sets having subsets already separated in the same execution of Algorithm \ref{alg:pigeon}. This means that the satisfaction of a given pigeonhole cut is not implied by the satisfaction of any previously separated cut, and thus redundant inequalities of this type are avoided.

    \begin{algorithm}[ht]
    	\SetKwInOut{Input}{\sc Input}\SetKwInOut{Output}{\sc Output}
    	\For{$i=1,\dots,n$}{
			$\bar V = \{i\}$;\\[-0.7ex]
			$\delta = 0$;\\[-0.7ex]
			\While{$|\bar V| < n$}{
				$j^* = \mbox{argmin}_{k \notin \bar V}~\left\{\zeta(k) = \sum_{i,j \in \bar V\cup\{k\}; i < j} s_{ij}\right\}$;\\[-0.7ex]
				$\bar V = \bar V \cup \{j^*\}$;\\[-0.7ex]
				\If{$\zeta(j^*)+ \delta +\epsilon < \omega(|\bar V|,K)$}{
					Add constraint \eqref{eq:mkpmlr:c10} for the current set $\bar V$;\\[-0.7ex]
					$\delta =\omega(|\bar V|,K) - \zeta(j^*)$
				}
			}
    	}
    	\caption{Pseudo-code for separating Pigeonhole cuts}\label{alg:pigeon}
    \end{algorithm}

\subsection{Branch-and-cut implementation}\label{subsec:branch-and-cut}

To solve model \eqref{eq:mkpmlr:fo}--\eqref{eq:mkpmlr:c10},
we implemented the two-phase method shown in Algorithm \ref{alg:bnc}. The first phase (steps 1--10 of the algorithm) is aimed at producing a strong initial model, which is then handed to the second phase (steps 11--12), which is responsible for solving the model via a standard branch-and-cut procedure. More specifically, we first find a heuristic solution of value UB by invoking the metaheuristic that is described in Section \ref{sec:heuristic}. Then, we iteratively solve the continuous relaxation of the model, separate violated inequalities described in Section \ref{subsec:valid-inequalities}, add them to the model, remove the previously added cuts that are not active in the current solution, and remove variables whose reduced cost $\bar c$ is larger than the gap between UB and the current continuous relaxation solution cost, $cost(curr)$. The first phase ends when there are no more cuts to be added. The resulting model typically has a reduced number of variables and a good number of strengthening cuts. A standard branch-and-cut method is then used to solve this model, with separation procedures invoked by means of callback functions at each node of the enumeration tree explored by the MILP solver. 
\begin{algorithm}[ht]
    	\SetKwInOut{Input}{\sc Input}\SetKwInOut{Output}{\sc Output}
    	$UB = ILS()$;\\[-0.7ex]
    	$curr = solveContinuousRelaxation(wkTRPM)$;\\[-0.7ex]
    	$cuts = separate(curr)$;\\[-0.7ex]
    	\Repeat{$sep = \{\}$}{
    	    $curr = solveContinuousRelaxation(wkTRPM)$;\\[-0.7ex]
    	    remove from $cuts$ and $wkTRPM$ all cuts that are non active in $curr$;\\[-0.7ex]
    	    remove from $wkTRPM$ all variables whose reduced cost $\bar c$ satisfies $\bar c+cost(curr) > UB$;\\[-0.7ex] 
    	    $sep = separate(curr)$;\\[-0.7ex]
    	    $cuts = cuts + sep$\;
    	}
    	Solver.setCallback(separate());\\[-0.7ex]
    	Solver.MIPSolve(wkTRPM)
    	\caption{Branch-and-cut algorithm.}\label{alg:bnc}
    \end{algorithm}
    
\section{Iterated Local Search}\label{sec:heuristic}
    
To obtain good-quality solutions in short computing times, we developed an \emph{iterated local search} (ILS). ILS is a simple yet powerful metaheuristic that obtained relevant results on a large number of combinatorial problems. We refer the interested reader to the recent survey by \cite{Lourenco2019}.
The ILS used in this work is shown in Algorithm \ref{alg:ils} and can be decomposed into three main components: construction phase, local search, and perturbation. In the construction phase (steps 1--4 of the algorithm), an initial solution is generated by a round robin method that distributes the clients randomly among the $K$ routes. Then, the main ILS loop (steps 5--20) iteratively invokes two local search procedures to improve the quality of the current solution, possibly updates current and best solutions, and then perturb the current solution to create a starting point for the next round of local search. 

More specifically, procedure \textit{relocate} randomly selects a client, removes them from their current position in the solution received as input, and relocates them in a lowest-cost position. It repeats the process until an improvement has been found or all clients have been considered.
Procedure $2opt^*$ is invoked with probability of 50\% or if \textit{relocate} fails in finding an improved solution. The $2opt^*$ procedure attempts to improve the current solution by replacing a pair of arcs $(i, i^+)$ and $(j, j^+)$ belonging to two different routes in the solution, with the pair $(i, j^+)$ and $(j, i^+)$. The process is repeated until an improvement is found or all pairs of arcs have been scanned. Both procedures return \emph{true} as soon as an improvement, if any, is found, and \emph{false} otherwise. 

When no improvement is found, the current solution is a local optimum whose value is compared to that of the incumbent solution. Then, if the current solution value is within a given range of the incumbent solution value, the current solution is maintained as starting point for a perturbation, otherwise it is replaced by the incumbent.
To escape from local optima, we adopted a perturbation procedure, called $pRelocate$, which randomly removes $p$ clients from the current solution at once, and reinserts them one at a time in lowest-cost positions. Parameter $p$ is initialized with the average number of clients per route ($\lfloor N/K \rfloor$) and incremented, up to $N-2$, whenever the current solution is identical to the solution obtained at the end of the previous iteration of the main ILS loop.

\begin{algorithm}[ht]
$curr = buildInitialSolution()$;\\[-0.7ex]
$best = curr$;\\[-0.7ex]
$p = N/K$;\\[-0.7ex]
$lastCurr = Null$;\\[-0.7ex]
\Repeat{stopping criterion}{
    \Repeat{imp = false}{
        $imp = relocate(curr)$;\\[-0.7ex]
        \If{imp = false \emph{\bf or} randomBoolean()$\leq0.5$}{
            $imp = 2opt^*(curr)$\;
        }
    }
    \eIf{$cost(curr) < cost(best)$}{
        $best = curr$;\\[-0.7ex]
    }{
        \If{$cost(curr)\cdot0.95> cost(best)$}{
            $curr = best$;\\[-1.4ex]
        }
    }
    \If{$lastCurr==curr$ \emph{\bf and} $p<N-2$}{
        $p = p + 1$\;
    }
    $lastCurr = curr$;\\[-0.7ex]
    $pRelocate(curr,p)$\;
}
\Return{best}
\caption{Pseudocode of the ILS procedure.}\label{alg:ils}
\end{algorithm}
              
\section{Computational Results}\label{sec:results}
In this section, we report on the outcome of extensive computational tests that we performed both on the real-world instances from our case study and on benchmark instances from the literature. The experiments aim at showing the behavior of the proposed methods under different problem components and instance characteristics, thereby assessing their performance in comparison to the existing literature. Our algorithms were coded in Java 8 and executed on an Intel i7-3820 processor with 8 cores running at 3.60GHz and 16GB of RAM, operating under \textit{Ubuntu 18.04}. As a MILP solver we adopted Gurobi 8.0. We switched on the \textit{intensive presolve} of Gurobi in order to attempt the elimination of redundant variables, while all other solver parameters were kept at their default values.
Unless stated otherwise, the branch-and-cut that we implemented (Algorithm \ref{alg:bnc}), referred to simply as \textit{B\&Cut} in the following, was allowed to run for at most two hours on each instance, whereas the ILS (Algorithm \ref{alg:ils}) was halted after 10,000 iterations. To perform some specific evaluations on the robustness of the ILS, or to compare its results with those from the literature, we also performed tests where the ILS was run 10 times on each instance and halted after a certain time limit, as outlined next in the specific sections.

\subsection{Benchmark Instances}
We first considered nine instances from our real-world case study. The results that we obtained are presented in Section \ref{subsec:real-results} and compared with the solutions adopted by the company. As the real-world instances are quite small, we also considered a number of larger benchmark instances from the literature, and solved them under different variants. In particular, we adopted the benchmark sets proposed by \cite{Luo2014} (denoted as \textit{LQL-instances} in the following), \cite{Christofides1969} (\textit{E-instances}),  \cite{Augerat98} (\textit{P-instances}), \cite{Christofides1979}  (\textit{CMT-instances}) and \cite{Ribeiro2012} (\textit{Rig-instances}).

The set of LQL-instances was originally proposed for the \textit{$k$-Travelling Repairman Problem with Distance Constraints} (kTRPDC), which is the kTRP variant in which each route cannot exceed a maximum input distance $D$. The set is composed of 180 instances derived from 6 TSPLIB instances (\cite{Reinelt1991}), that contain 30, 40 and 50 vertices to be visited by 6, 8 and 10 repairmen, respectively. The set has been made available by \cite{Luo2014} at \url{www.computational-logistics.org/orlib/mtrpd}.
To adapt our methods to the kTRPDC, we simply set all client weights to one and all repair times to zero. The maximum distance constraint was easily taken into account in the ILS, and was added to our MILP model as the following lazy constraints
\begin{align}
& \sum_{i=0}^{n} \sum_{j=1, i\neq j}^{n}  \left( c_{ij}+r_j \right)   f^k_{ij} \le D & \forall k = 1,\ldots,n.  \label{eq:DC}
\end{align}
It is worth noting that constraints \eqref{eq:DC} were never activated on any LQL-instance. The kTRPDC results that we obtained are shown in Section \ref{subsec:kTRPDC} and compared with \cite{Luo2014}.

The E, P, and CMT instances were originally proposed for the Capacitated Vehicle Routing Problem. E-instances contain from 22 to 101 vertices, P-instances from 16 to 76 vertices and CMT-instances from 51 to 200 vertices. Following \cite{Ngueveu2010}, who used these instances to test their kTRP algorithms, we relaxed the capacity constraint and supposed each client has weight equal to one and null service time. The results that we obtained on these kTRP instances are also presented in Section \ref{subsec:kTRPDC}.

The Rig-instances were proposed by \cite{Ribeiro2012} as WRRP test cases. They combine rigs of 5 and 10 vertices, with  sets of 50, 100 and 500 clients. 
The particularity of Rig-instances is the presence of $\Delta$ rigs, which can be interpreted as multiple depots for the wkTRP. We considered this feature by adding to our wkTRP graphs $\Delta+1$ artificial vertices, $v_0, v_1, \ldots, v_{\Delta}$, having zero weight and zero service time. Vertex $v_0$ is our unique depot, whereas vertices $v_1, v_2, \ldots, v_{\Delta}$ are the rigs' starting points. 
A new time matrix $c'$ is then computed, starting from the original $c$ matrix, by setting $c'_{ij}$ equal to: 0 if $i = 0$ and $j \le \Delta$; 0 if $i > \Delta$ and $j = 0$; $t_{ij}$ if $i \ge 1$ and $j > \Delta$;
+$\infty$ if $i = 0$ and $j > \Delta$; 
+$\infty$ if $i \le \Delta$ and $j \le \Delta$; 
+$\infty$ if $i > \Delta$ and $1 \le j \le \Delta$.

The WRRP also imposes a deadline $D_l$ on each client $l= \Delta+1,\ldots , n$. We easily included this feature in our ILS, and modeled it in our MILP model by introducing the following lazy constraints
\begin{align}
& \sum_{i=0}^{n} \sum_{j=1, i\neq j}^{n}  \left( c_{ij}+r_j \right)   f^k_{ij} \le D_l & \forall l = \Delta+1,\ldots,n.  \label{eq:DCk} 
\end{align}
The MILP model was run only on instances with 50 clients, as these already proved to be very challenging. Also in this test, as happened for constraints \ref{eq:DC},  we detected that constraints \eqref{eq:DCk} were never activated on any of the Rig-instances.

To perform a sounded test of our methods on the wkTRP, we produced a weighted version of the LQL-instances, called \textit{wLQL} in the following, by randomly assigning with uniform distribution a weight between 0.5 and 2.0 to each client. In this way, we obtained a proper wkTRP set, that we used to computationally test our methods in Section \ref{subsec:wkTRP}.

All the benchmark instances that we used and the complete computational results that we obtained are available  on-line at \url{https://github.com/spideryzarc/wkTRP}.   
       
\subsection{Results on the Real-World Instances}\label{subsec:real-results}

We tested nine instances based on the activities performed by our case study company in nine ordinary working days in the city of Rio de Janeiro. Each instance is composed of the list of defective devices, their locations, their weights (expressed in fine costs per minute) and their repair times, as well as the traveling times required to move from one location to another. The traveling times were obtained by using Google Map's Web service. The repair times and profits were computed by taking into account the nature of the malfunction and the device and considering average values of the historical data at the company. 

\begin{table}[ht]
\caption{Computational results on Rio de Janeiro instances and comparison with company solutions}\label{tabRioSimpl}
\centering 
\small
%\resizebox{\textwidth}{!}{%
\begin{tabular}{lrr rrr HrrrHHHH rr}
\toprule
		\multicolumn{3}{c}{instance} & \multicolumn{3}{c}{ILS} & \multicolumn{8}{c}{B\&C} & \multicolumn{2}{c}{company}\\ 
\cmidrule(lr){1-3} \cmidrule(lr){4-6}  \cmidrule(lr){7-14} \cmidrule(lr){15-16}
name{} & $n$ & $k$ & {cost} & {cost*} & {time} & {RTt} & {cost} & {nodes} & {time} & {SEPt} & {Pgn} & {$z$Act} & {$f$Act}& {cost*}& gap\%\\ 
\cmidrule(lr){1-3} \cmidrule(lr){4-6}  \cmidrule(lr){7-14} \cmidrule(lr){15-16}
RIO\_01\_08 & 12 & 2 & 971.85  &1155.22 & 0.09 & 0.35 & 971.85  & 459    & 9.02  & 0.05 & 216    & 506    & 365    & 1728.57 &  33.17 \\
RIO\_07\_08 & 14 & 2 & 1082.43 &1109.65 & 0.07 & 0.28 & 1082.43 & 624    & 9.38  & 0.05 & 399    & 508    & 332    & 1815.62 &  38.88 \\
RIO\_08\_08 & 18 & 2 & 1641.18 &1756.57 & 0.11 & 1.96 & 1641.18 & 618    & 37.82 & 0.12 & 1024   & 1469   & 1751   & 4063.72 &  56.77 \\
RIO\_10\_08 & 17 & 2 & 1433.61 &1605.10 & 0.09 & 1.41 & 1433.61 & 662    & 28.29 & 0.09 & 802    & 915    & 781    & 3217.05 &  50.11 \\
RIO\_02\_10 & 13 & 1 & 1843.73 &1622.85 & 0.03 & 0.08 & 1843.73 & 0      & 0.11  & 0    & 0      & 121    & 110    & 2540.67 &  36.13 \\
RIO\_03\_10 & 16 & 2 & 1529.24 &2473.11 & 0.08 & 0.65 & 1529.24 & 0      & 0.74  & 0    & 420    & 413    & 440    & 4254.67 &  41.87 \\
RIO\_09\_10 & 16 & 2 & 1356.93 &1613.26 & 0.08 & 0.83 & 1356.93 & 1      & 3.94  & 0    & 485    & 635    & 599    & 1923.80 &  16.14 \\
RIO\_10\_10 & 11 & 1 & 1190.96 &1155.56 & 0.03 & 0.06 & 1190.96 & 0      & 0.09  & 0    & 0      & 129    & 145    & 1355.64 &  14.76 \\
RIO\_11\_10 & 18 & 2 & 1672.57 &1884.52 & 0.11 & 1.74 & 1672.57 & 920    & 39.06 & 0.13 & 1024   & 1590   & 1830   & 2079.79 &  9.39  \\
\cmidrule(lr){1-3} \cmidrule(lr){4-6}  \cmidrule(lr){7-14} \cmidrule(lr){15-16}
averages      &    &   & 1413.61 &1597.32 & 0.08 & 0.82 & 1413.61 & 364.89 & 14.27 & 0.05 & 485.56 & 698.44 & 705.89 & 2553.28 &  33.02 \\
\bottomrule
\end{tabular}
%}

\end{table}

The results that we obtained are provided in Table \ref{tabRioSimpl}. Column $n$ gives the number of devices and column $K$ the number of repairmen. For the ILS, we report the solution cost (cost) and the run time in seconds (time). The cost was computed by using the estimated average service times. Column {cost*} gives, instead, the value of the best solution found by the ILS recalculated by replacing the average repair times with the real ones incurred in those events. 
For the B\&Cut, we report the cost of the solution obtained, the number of nodes explored and the computational time in seconds. We do not report the recalculated cost as it was always equal to the one shown for the ILS.
In the penultimate column, we report the cost of the solution that was manually created and adopted by the company, that we evaluated by using our average weights, our traveling times from Google Map, and the real repair times. Then, in column {gap\%}, we report the percentage gap between the manual and the ILS solution costs, computed as (Company cost* - ILS cost*)/Company cost*$\times 100\%$. The last line reports average values for the nine instances.

The instances from this case study may be considered quite small, especially when compared to the benchmark sets from the literature that are addressed in the next sections. Nevertheless, they are useful to highlight some interesting facts. The monetary saving obtained by the ILS with respect to the manual company solutions is substantial, achieving an average gap of about 33\%, with peaks that are above 50\%. Such values were obtained very quickly, always in less than a second. This emphasizes the relevance of using an optimization tool for the wkTRP in the real-world application.
The B\&Cut always achieved the same solution values obtained by the ILS, in addition to proving their optimality, in less than 15 seconds on average and never required more than 40 seconds. 

\subsection{Results and Comparative Analysis on the kTRPDC and the kTRP}\label{subsec:kTRPDC}

In Table \ref{tabTSPLibAggregate}, we report the results that we obtained by adapting our methods to the kTRPDC and testing them on the LQL set. The columns reported for ILS and B\&Cut have the same meanings than those reported in Table \ref{tabRioSimpl}, but now each line reports average values on 10 instances sharing the same seminal routing instance (shown in column \textit{name}) and the same value of $n$. 

\begin{table}[tbh]
\caption{Computational results for the kTRPDC on the LQL set (10  instances per line) and comparison with \cite{Luo2014} and \cite{nucamendi2016} (* = one instance unsolved to proven optimality)} \label{tabTSPLibAggregate}
% \resizebox{\textwidth}{!}{%
%\scriptsize
\small
\centering
	\begin{tabular}{llrrHrrrHHHHrrr}
\toprule
\multicolumn{2}{c}{instance} & \multicolumn{2}{c}{ILS} & \multicolumn{8}{c}{B\&Cut} & \multicolumn{1}{c}{Luo BPC1} & \multicolumn{1}{c}{Luo BPC2} & \multicolumn{1}{c}{Nucamendi} \\ 
\cmidrule(lr){1-2} \cmidrule(lr){3-5} \cmidrule(lr){6-12} \cmidrule(lr){13-13}
\cmidrule(lr){14-14} \cmidrule(lr){15-15}
{name} & {$n$} & {cost} & {time} & {RTt} & {cost} & {nodes} & {time} & {SEPt} & {Pgn} & {$z$Act} & {$f$Act} & time & time & time\\
\cmidrule(lr){1-2} \cmidrule(lr){3-5} \cmidrule(lr){6-12} \cmidrule(lr){13-13}
\cmidrule(lr){14-14} \cmidrule(lr){15-15}
brd14051 & 30 & 85093.54  & 0.27 & 1.40  & 85093.54  & 13.10  & 9.41   & 0.04 & 1143.50  & 332.40  & 262.20  &   15.80   & 8.80 & 4.66   \\
 d15112   & 30 & 229130.79 & 0.29 & 1.45  & 229130.79 & 28.20  & 3.67   & 0.03 & 1243.40  & 325.20  & 266.80  &   12.80   & 25.20 & 9.58  \\
 d18512   & 30 & 84818.95  & 0.26 & 1.36  & 84818.95  & 0.60   & 2.64   & 0.01 & 1227.00  & 342.60  & 263.40  &   9.80    & 11.60 & 6.85 \\
 fnl4461  & 30 & 48287.50  & 0.26 & 1.19  & 48287.50  & 8.30   & 3.65   & 0.02 & 1123.60  & 276.60  & 210.40  &  10.70   & 6.30 & 2.83    \\
 nrw1379  & 30 & 29994.97  & 0.27 & 1.45  & 29994.97  & 31.60  & 9.90   & 0.07 & 1204.10  & 333.60  & 292.10  &   15.40   & 21.80 & 5.83  \\
 pr1002   & 30 & 164763.90 & 0.26 & 1.47  & 164763.90 & 18.70  & 4.76   & 0.03 & 1216.60  & 402.50  & 294.60  &   13.20   & 12.20 & 8.09  \\
 brd14051 & 40 & 106145.48 & 0.42 & 4.69  & 106145.48 & 186.30 & 43.49  & 0.41 & 3521.70  & 811.50  & 707.70  &  194.70  & 83.20 & 30.37  \\
 d15112   & 40 & 290748.64 & 0.47 & 3.80  & 290748.64 & 0.60   & 4.88   & 0.01 & 2342.10  & 440.60  & 340.00  &  34.80   & 27.80 & 24.96  \\
 d18512   & 40 & 107813.36 & 0.40 & 3.95  & 107813.36 & 7.50   & 8.21   & 0.05 & 2673.70  & 552.30  & 403.60  & 47.20   & 24.20 & 11.74   \\
 fnl4461  & 40 & 62672.49  & 0.45 & 4.44  & 62672.49  & 66.50  & 21.92  & 0.22 & 2730.10  & 601.00  & 475.10  &  300.00  & 73.40 & 17.21  \\
 nrw1379  & 40 & 36379.77  & 0.45 & 3.62  & 36379.77  & 17.00  & 12.54  & 0.09 & 2431.50  & 402.40  & 342.20  &  64.00   & 66.60 & 20.66  \\
 pr1002   & 40 & 217071.57 & 0.41 & 4.34  & 217071.57 & 196.00 & 41.26  & 0.43 & 3173.30  & 892.70  & 599.00  &  389.00  & 209.60 & 52.12 \\
 brd14051 & 50 & 127480.58 & 0.67 & 10.28 & 127480.58 & 72.30  & 73.29  & 0.54 & 4681.20  & 649.50  & 582.00  &  1483.90 & 259.10 & 51.54 \\
 d15112   & 50 & 356433.93 & 0.70 & 8.85  & 356433.93 & 67.40  & 33.37  & 0.49 & 5337.60  & 675.50  & 574.80  &  569.40  & 78.70 & 53.75  \\
 d18512   & 50 & 132115.39 & 0.66 & 11.32 & 132115.39 & 210.50 & 49.13  & 0.93 & 5766.40  & 999.80  & 723.30  &  341.10  & 213.10 & 75.64 \\
 fnl4461  & 50 & 76642.75  & 0.67 & 9.02  & 76642.75  & 0.80   & 20.14  & 0.13 & 4263.40  & 586.00  & 491.40  & 141.10  & 53.20 & 32.11  \\
 nrw1379  & 50 & 45169.23  & 0.66 & 10.95 & 45169.23  & 898.60 & 189.44 & 5.34 & 18354.60 & 2515.30 & 2399.60 &  1290.20$^*$\hspace{-1ex} & 1184.20$^*$\hspace{-1ex} & 79.66 \\
 pr1002   & 50 & 268605.26 & 0.66 & 10.52 & 268605.26 & 59.30  & 36.73  & 0.38 & 4968.80  & 731.10  & 552.00  & 288.70  & 94.50 & 60.70  \\ 
\cmidrule(lr){1-2} \cmidrule(lr){3-5} \cmidrule(lr){6-12} \cmidrule(lr){13-13}
\cmidrule(lr){14-14} \cmidrule(lr){15-15}
\multicolumn{2}{l}{averages} & 137187.12 & 0.46 & 5.23  & 137187.12 & 104.63 & 31.58  & 0.51 & 3744.59  & 659.48  & 543.34  & 290.10  & 136.31 & 30.46  \\
 \bottomrule
%\multicolumn{15}{l}{* = one instance unsolved to proven optimality}
\end{tabular}%
%}
\end{table}

We also selected the three best methods currently available in the literature for the problem, namely, the two branch-and-price-and-cut methods developed for the kTRPDC by \cite{Luo2014} (BPC1 and BPC2) and the combination of MILP model and metaheuristic procedure proposed by \cite{nucamendi2016} for the kTRP. For these three methods we only report the average solution times.
It is worth noting that, as also stressed by \cite{nucamendi2016}, the distance constraint is never activated in any of the optimal solutions of the LQL benchmark set, so the results obtained by kTRPDC methods (ILS, B\&Cut, BPC1 and BPC2) are directly comparable with those obtained by a kTRP method (Nucamendi). 
The BPC1 and BPC2 algorithms were executed on a PC with 2.26Ghz CPU by using ILOG CPLEX 12.0 as MILP solver and imposing a maximum run time of 10800 CPU seconds. The Nucamendi algorithm was instead run on a PC with 3 GHz CPU by using  ILOG CPLEX 12.4. 
Both our B\&Cut and the Nucamendi algorithm solved all instances to proven optimality. Instead, an instance of the group \texttt{nrw1379\_1} with 50 vertices was not solved to proven optimality by neither BPC1 nor BPC2 within the 3 hours of time limit that they were allowed. 
The average CPU times required by the B\&Cut were 5.67, 22.05 and 67.02 seconds for instances having 30, 40 and 50 vertices, respectively, and 31.58 seconds on average for the entire LQL set. The longest solution time was 1064.14 seconds and was required for an instance of the group \texttt{nrw1379-1} having 50 vertices. Although enumeration methods are employed for cuts separation, the total CPU spent in the separation process was very small when compared to the overall running time. Overall, the performance of the B\&Cut can be considered superior to that of BPC1 and BPC2, and comparable to that of Nucamendi.
Even if it was originally implemented to solve a different problem (the wkTRP), the ILS managed to reach all proven optimal solutions in very quick times, never requiring more than a second and on average slightly less than half a second. This proves both the very high effectiveness and speed of the metaheuristic. 
\begin{table}[t]
\caption{Computational results for the kTRP on E- and P-instances and comparison with \cite{nucamendi2016} (t.lim. = 7200 seconds, * = averages on solutions solved by Nucamendi))}\label{tabEPComp}
\centering
{\small
%\resizebox{\textwidth}{!}{%
\begin{tabular}{l rr rrrr rr}
\toprule
\multicolumn{1}{l}{instance} & \multicolumn{2}{c}{ILS} & \multicolumn{4}{c}{B\&Cut} & \multicolumn{2}{c}{{Nuncamendi}} \\ 
\cmidrule(lr){1-1} \cmidrule(lr){2-3} \cmidrule(lr){4-7} \cmidrule(lr){8-9}
{name}  & {cost} & {time} &{cost} & {nodes} & {time} & {gap\%} & {time} & {gap\%} \\ 
\cmidrule(lr){1-1} \cmidrule(lr){2-3} \cmidrule(lr){4-7} \cmidrule(lr){8-9}
E-n22-k4 	&	819.39	&	0.15	&	819.39	&	1	&	4.32	&	0.00	&	4.90	&	0.00	\\
E-n23-k3 	&	1555.87	&	0.16	&	1555.87	&	0	&	1.37	&	0.00	&	9.23	&	0.00	\\
E-n30-k3 	&	1871.08	&	0.16	&	1871.08	&	3744	&	1100.84	&	0.00	&	119.11	&	0.00	\\
E-n30-k4 	&	1643.30	&	0.22	&	1643.30	&	3046	&	301.07	&	0.00	&	22.95	&	0.00	\\
E-n33-k4 	&	2819.43	&	0.26	&	2819.43	&	1399	&	414.74	&	0.00	&	24.29	&	0.00	\\
E-n51-k5 	&	2209.64	&	0.36	&	2209.64	&	1697	&	t.lim.	&	1.15	&	2347.51	&	0.00	\\
E-n76-k7 	&	2945.25	&	0.68	&	2945.25	&	46	&	t.lim.	&	2.23	&	 --                  	&	 --           	\\
E-n76-k8 	&	2677.39	&	0.71	&	2677.39	&	275	&	t.lim.	&	1.72	&	 --                  	&	 --           	\\
E-n76-k10 	&	2310.09	&	0.83	&	2310.09	&	1218	&	2797.17	&	0.00	&	1700.64	&	0.00	\\
E-n76-k14 	&	2005.40	&	1.17	&	2005.40	&	1	&	271.04	&	0.00	&	236.64	&	0.00	\\
E-n76-k15 	&	1962.47	&	1.28	&	1962.47	&	1	&	109.30	&	0.00	&	105.43	&	0.00	\\
E-n101-k14 	&	2922.82	&	1.78	&	2922.82	&	73	&	t.lim.	&	0.55	&	 --                  	&	 --           	\\
\cmidrule(lr){1-1} \cmidrule(lr){2-3} \cmidrule(lr){4-7} \cmidrule(lr){8-9}
averages*	&	1910.74	&	0.51	&	1910.74	&	1234.11	&	1355.59	&	0.13	&	507.86	&	0.00	\\
averages 	&	2145.18	&	0.65	&	2145.18	&	958.42	&	2817.17	&	0.47	&		&		\\
\cmidrule(lr){1-1} \cmidrule(lr){2-3} \cmidrule(lr){4-7} \cmidrule(lr){8-9}
P-n16-k8 	&	382.90	&	0.09	&	382.90	&	0	&	0.36	&	0.00	&	1.90	&	0.00	\\
P-n19-k2 	&	812.15	&	0.09	&	812.15	&	0	&	1.27	&	0.00	&	9.13	&	0.00	\\
P-n20-k2 	&	905.19	&	0.09	&	905.19	&	1	&	2.07	&	0.00	&	11.60	&	0.00	\\
P-n21-k2 	&	937.10	&	0.11	&	937.10	&	1	&	6.36	&	0.00	&	11.07	&	0.00	\\
P-n22-k2 	&	993.10	&	0.13	&	993.10	&	0	&	1.87	&	0.00	&	11.00	&	0.00	\\
P-n22-k8 	&	623.40	&	0.18	&	623.40	&	0	&	0.29	&	0.00	&	3.08	&	0.00	\\
P-n23-k8 	&	561.33	&	0.18	&	561.33	&	0	&	0.29	&	0.00	&	2.68	&	0.00	\\
P-n40-k5 	&	1537.79	&	0.27	&	1537.79	&	797	&	196.62	&	0.00	&	213.28	&	0.00	\\
P-n45-k5 	&	1912.31	&	0.31	&	1912.31	&	3505	&	3179.19	&	0.00	&	495.82	&	0.00	\\
P-n50-k7 	&	1547.89	&	0.43	&	1547.89	&	288	&	211.15	&	0.00	&	117.48	&	0.00	\\
P-n50-k8 	&	1448.92	&	0.49	&	1448.92	&	2164	&	958.75	&	0.00	&	185.76	&	0.00	\\
P-n50-k10 	&	1296.48	&	0.57	&	1296.48	&	1	&	13.57	&	0.00	&	112.32	&	0.00	\\
P-n51-k10 	&	1419.43	&	0.57	&	1419.43	&	75	&	82.91	&	0.00	&	84.17	&	0.00	\\
P-n55-k7 	&	1766.56	&	0.49	&	1766.56	&	3386	&	3033.76	&	0.00	&	790.31	&	0.00	\\
P-n55-k8 	&	1614.61	&	0.48	&	1614.61	&	1	&	136.19	&	0.00	&	170.89	&	0.00	\\
P-n55-k10 	&	1438.60	&	0.67	&	1438.60	&	0	&	15.53	&	0.00	&	117.69	&	0.00	\\
P-n55-k15 	&	1280.92	&	0.81	&	1280.92	&	0	&	4.23	&	0.00	&	48.01	&	0.00	\\
P-n60-k10 	&	1676.35	&	0.72	&	1676.35	&	1245	&	674.62	&	0.00	&	620.81	&	0.00	\\
P-n60-k15 	&	1462.50	&	0.86	&	1462.50	&	1	&	13.14	&	0.00	&	 --                  	&	 --           	\\
P-n65-k10 	&	1928.46	&	0.76	&	1928.46	&	4239	&	6762.68	&	0.00	&	915.83	&	0.00	\\
P-n70-k10 	&	2097.17	&	0.82	&	2097.17	&	2421	&	5877.93	&	0.00	&	1415.78	&	0.00	\\
P-n76-k4 	&	4673.05	&	0.50	&	4673.05	&	6	&	t.lim.	&	3.45	&	t.lim.	&	8.17	\\
P-n76-k5 	&	3820.02	&	0.54	&	3820.02	&	19	&	t.lim.	&	2.50	&	t.lim.	&	7.29	\\
\cmidrule(lr){1-1} \cmidrule(lr){2-3} \cmidrule(lr){4-7} \cmidrule(lr){8-9}
averages*	&	1576.08	&	0.42	&	1576.08	&	824.95	&	1617.14	&	0.27	&	897.21	&	0.70	\\
averages 	&	1571.14	&	0.44	&	1571.14	&	789.13	&	1547.40	&	0.26	&		&		\\
\bottomrule
\end{tabular}%
%}

 }
\end{table}

We performed two additional tests on the kTRP (thus disregarding the distance constraint) in order to obtain a better comparison of our methods with respect to those in \cite{nucamendi2016}. The first such test is presented in
Table \ref{tabEPComp}, where we show the results obtained by our ILS and B\&Cut on the E- and {P-instances}. For the approach from the literature, we show the execution time in seconds and the final percentage gap it produced. For the B\&Cut, we report cost, number of nodes, execution time and percentage gap.

The B\&Cut could solve most of the instances within the two-hour time limit. For the five instances that were not solved to optimality (En51k5, En76k7, En76k8, Pn76k4 and Pn76k5) it produced low gaps below 4\%. 
The ILS heuristic always {achieved the best solution value requiring} at most 1.78 seconds. The method by \cite{nucamendi2016}, that we recall has been specifically developed to solve the kTRP, has a better performance on the E-instances with respect to both gap and time. On the P-instances, however, it is faster but produces a larger average gap. This can be imputed to the two largest instances, P-n76-k4 and P-n76-k5. The ILS is still very effective, as it finds the best solution value on each instance {in less than half a second on average}.

The second additional test that we performed aimed at assessing the performance of the ILS on the large-size CMT instances, and it is shown in Table \ref{tabHeuCMT}.
\cite{nucamendi2016} proposed an {Iterated Greedy (IG) metaheuristic} that was run once on each instance. To test our ILS, we executed it 10 times. We report  minimum, average and maximum solution costs found in the 10 runs. We attempted two time limits, just one second and ten seconds, so as to gain some insight in the convergence of the algorithm. The IG requires about one minute on average and slightly more than five minutes in the worst case (instance CMT5). The ILS was much faster due to the strict time limits that we imposed. Nevertheless, it obtained results that are comparable for the one-second time limit, and even slightly better for the ten-second time limit. The very limited difference between minimum and maximum costs further assess the robustness of the algorithm.
We also solved the CMT-instances with our B\&Cut method. For the sake of conciseness, we avoid presenting full computational results, but we mention that only instances CMT2 and CMT7 were solved to optimality within the two-hour time limit and that the B\&Cut could not improve the ILS solution on any instance.

\begin{table}[htb]
\caption{Comparison of heuristics for the kTRP on CMT-instances (* = on solutions solved by Nucamendi)}\label{tabHeuCMT}
\centering
{\small %\scriptsize
%\resizebox{\textwidth}{!}{%
\begin{tabular}{l rrrrrr rr}
\toprule
    &  \multicolumn{6}{c}{{this work}}  & \multicolumn{2}{c}{{Nucamendi}}  \\
	\cmidrule(lr){2-7} \cmidrule(lr){8-9} 
	{instance}  &  \multicolumn{3}{c}{{ILS (1 sec. t.lim.)}}  & \multicolumn{3}{c}{{ILS (10 sec. t.lim.)}}  &\multicolumn{2}{c}{{IG}}  \\
	\cmidrule(lr){1-1} \cmidrule(lr){2-4} \cmidrule(lr){5-7}\cmidrule(lr){8-9}
	name & {min cost} & {avg cost} & {max cost} & {min cost} & {avg cost} & {max cost} & {cost} &  {time} \\ 	\cmidrule(lr){1-1} \cmidrule(lr){2-4} \cmidrule(lr){5-7}\cmidrule(lr){8-9}
    CMT1                                  & 2209.64      & 2212.56      & 2223.78      & 2209.64      & 2209.64       & 2209.64      & 2209.64 & 0.70 \\
	CMT2                                  & 2310.09      & 2310.09       & 2310.09      & 2310.09      & 2310.09       & 2310.09      & 2310.09 & 4.20 \\
	CMT3                                  & 4002.90       & 4005.12       & 4010.30       & 4002.90       & 4005.12       & 4010.30       & 4002.90 &  14.94\\
	CMT4                                  & 4953.94      & 4957.38      & 4970.55      & 4953.94      & 4953.94       & 4953.94      & 4953.94 & 80.95 \\
	CMT5                                  & 5745.94      & 5760.80      & 5780.24      & 5745.94      & 5747.32      & 5748.62      & 5748.36& 313.67 \\
	CMT6                                  & 1921.50       & 1921.50        & 1921.50       & 1921.50       & 1921.50        & 1921.50       & -- & -- \\
	CMT7                                  & 2202.30       & 2202.30        & 2202.30       & 2202.30       & 2202.30        & 2202.30       & -- & -- \\
	CMT8                                  & 3689.31      & 3689.31       & 3689.31      & 3689.31      & 3689.31       & 3689.31      & -- & -- \\
	CMT9                                  & 4547.52      & 4553.58      & 4560.54      & 4547.52      & 4548.12      & 4549.11      & -- & -- \\
	CMT10                                 & 5621.10       & 5627.13      & 5645.72      & 5620.08      & 5620.16       & 5620.79      & -- & -- \\
	CMT11                                 & 7201.30       & 7203.45      & 7206.40       & 7197.21      & 7198.53      & 7201.30       & 7200.68 & 27.10\\
	CMT12                                 & 3530.74      & 3532.78      & 3534.07      & 3530.24      & 3530.24       & 3530.24      & 3530.24 & 13.84 \\
	CMT13                                 & 6520.57      & 6520.57       & 6520.57      & 6520.57      & 6520.57       & 6520.57      & -- & -- \\
	CMT14                                 & 3422.56      & 3423.39      & 3425.22      & 3422.56      & 3422.56       & 3422.56      & -- & --  \\ \cmidrule(lr){1-1} \cmidrule(lr){2-4} \cmidrule(lr){5-7}\cmidrule(lr){8-9}
    {averages*}                                 & {4279.22}	& {4283.17}	&{4290.78}	&{4278.57}	&{4279.27}	&{4280.59}	&{4279.41}	&{65.06} \\
{averages}                                 & {4134.24}	& {4137.14}	& {4142.90}	& {4133.84}	&{4134.24}	& {4135.02}
      &  &   \\
      \bottomrule
%\multicolumn{9}{l}{* = on solutions solved by Nucamendi}
\end{tabular}
%}
}
\end{table}

\subsection{Results and Comparative Analysis on the WRRP}\label{subsec:WRRP}
For what concerns the WRRP, we tested our B\&Cut on Rig-instances with 50 clients, and our ILS on all Rig-instances with up to 500 clients. The outcome of the first test is presented in Table \ref{tabRig50}, and the outcome of the second one in Table \ref{tabHeuRigAggregate}, where we also compare the ILS with the metaheuritic methods by \cite{Ribeiro2012}.

\begin{table}[htb]
 \caption{Computational results for the WRRP on the Rig-instances (t.lim. = 7200 seconds)} \label{tabRig50}
\centering
{\small
 %\resizebox{\textwidth}{!}{%
\begin{tabular}{l rr Hrrrrrrrr}
\toprule
\multicolumn{1}{l}{instance} & \multicolumn{2}{c}{ILS} & \multicolumn{9}{c}{B\&Cut} \\ 
\cmidrule(lr){1-1} \cmidrule(lr){2-3} \cmidrule(lr){4-12}
{name}  & {cost} & {time} & {RTt} & {cost} & {gap\%}    & {nodes} & {time} & {sep.time} & {\#pgn} & {\#$z$act} & {\#$f$act} \\ 
\cmidrule(lr){1-1} \cmidrule(lr){2-3} \cmidrule(lr){4-12}
{50/5–1}                                & 42682.12           & 0.88               & 370.99                 & 42682.12            & 4.24            & 197            & t.lim.            & 3.32               & 20400           & 5676         & 5064          \\
{50/5–2}                               & 45516.36           & 0.89               & 311.18                 & 45516.36            & 3.26            & 963            & t.lim.            & 15.61              & 70312           & 16745        & 20678         \\
{50/5–3}                               & 41622.72           & 0.95               & 430.34                 & 41622.72            & 4.03            & 223            & t.lim.            & 4.08               & 26190           & 7275         & 6301          \\
{50/5–4}                           & 40524.17           & 0.82               & 281.93                 & 40524.17            & 0.00            & 1180           & 5398.13            & 8.42               & 32058           & 5031         & 3687          \\
{50/5–5}                              & 42785.74           & 1.03               & 370.07                 & 42785.74            & 3.20            & 713            & t.lim.            & 10.55              & 35683           & 8234         & 9801          \\
{50/5–6}                              & 44682.51           & 0.85               & 331.83                 & 44682.51            & 2.71            & 282            & t.lim.            & 5.59               & 27595           & 5544         & 4533          \\
{50/5–7}                              & 51554.44           & 0.80                & 169.51                 & 51554.44            & 0.00            & 113            & 491.02             & 0.73               & 13917           & 3258         & 2526          \\
{50/5–8}                              & 37665.35           & 0.77               & 246.88                 & 37665.35            & 2.11            & 708            & t.lim.            & 9.18               & 31003           & 4736         & 3232          \\
{50/5–9}                              & 43498.26           & 0.98               & 266.52                 & 43498.26            & 0.56            & 1288           & t.lim.            & 11.29              & 26722           & 4204         & 3332          \\
{50/5–10}                             & 42970.28           & 0.93               & 324                    & 42970.28            & 1.92            & 396            & t.lim.            & 6.37               & 29205           & 5620         & 4034          \\
\cmidrule(lr){1-1} \cmidrule(lr){2-3} \cmidrule(lr){4-12}
{averages}          & 43350.20 &	0.89	&310.33	&43350.20	&2.20	&606		& 6349.88	&7.51	&31308	&6632	&6318\\
\cmidrule(lr){1-1} \cmidrule(lr){2-3} \cmidrule(lr){4-12}
{50/10–1}                               & 30486.13           & 1.84               & 167.96                 & 30486.13            & 0.50            & 1277           & t.lim.            & 12.85              & 19383           & 2542         & 2263          \\
{50/10–2}                             & 30933.87           & 1.31               & 117.65                 & 30933.87            & 2.41            & 543            & t.lim.            & 11.02              & 18411           & 1837         & 1594          \\
{50/10–3}                             & 29352.94           & 1.60                & 145.08                 & 29352.94            & 0.00            & 1736           & 5088.07            & 11.92              & 18890           & 2536         & 1693          \\
{50/10–5}                             & 28826.23           & 1.43               & 149.35                 & 28826.23            & 1.32            & 1031           & t.lim.             & 13.58              & 22930           & 2385         & 1916          \\
{50/10–5}                             & 29348.40            & 1.46               & 195.64                 & 29348.40            & 2.03            & 1025           & t.lim.             & 17.04              & 27287           & 3064         & 2290          \\
{50/10–6}                         & 30462.58           & 1.71               & 95.36                  & 30462.58            & 0.00            & 1842           & 3781.27            & 16.80               & 36906           & 3115         & 1570          \\
{50/10–7}                         & 37502.23           & 1.56               & 103.97                 & 37502.23            & 0.00            & 910            & 808.93             & 5.27               & 11485           & 1830         & 1530          \\
{50/10–8}                          & 26451.65           & 1.60                & 130.86                 & 26451.65            & 1.53            & 1108           & t.lim.            & 14.21              & 18464           & 2341         & 1587          \\
{50/10–9}                          & 31350.39           & 1.64               & 108.71                 & 31350.39            & 0.00            & 1331           & 4589.38            & 12.99              & 20254           & 2050         & 1788          \\
{50/10–10}                        & 29825.70            & 1.69               & 144.43                 & 29825.70            & 1.15            & 1988           & t.lim.            & 20.75              & 14142           & 2180         & 1808          \\
\cmidrule(lr){1-1} \cmidrule(lr){2-3} \cmidrule(lr){4-12}
{averages}   &30454.01	&1.58	&135.90	&30454.01	&0.89	&1279	&5748	&13.64	&20815	&2388	&1803    \\ 
\bottomrule
\end{tabular}%
%}
 }
\end{table}    

Table \ref{tabRig50} presents some additional columns for the B\&Cut with respect to the previous tables: \textit{gap\%} gives the percentage gap provided by Gurobi at the end of the run, \textit{sep.time} gives the time spent in the separation procedures, whereas \textit{\#pgn}, \textit{\#$z$act} and \textit{\#$f$act}, give, respectively, the number of violated pigeon hole, $z$- and $f$-activation cuts found during the execution.
The B\&Cut could solve to proven optimality only 6 instances out of 20 within the time limit of 2 hours. However, the percentage gap for the remaining instances is quite low, never exceeding 5\%. Instances with 5 rigs (first block of lines) appear to be more difficult than those with 10 rigs (second block), with an average percentage gap of 2.2\% against 0.89\%. The number of explored nodes is around 600 for the first block, and more than double for the second block. The separation time appears to be negligible with respect to the execution time. Among the cuts, we notice that the number of violated pigeonhole cuts is around five times larger than that of $z$- and $f$-activation cuts.
Notably, the ILS has found the best solution value over all 20 instances, reaching it in very low time: less than a second on average for the first group, and less than two seconds for the second group.

In the work by \cite{Ribeiro2012}, three different metaheuristics were implemented for the WRRP, namely, an ILS, a Clustering Search (CS) and an Adaptive Large Neighborhood Search (ALNS). They were tested on a PC with 2GhZ on the Rig-instances with a time limit of 30 seconds and performing 10 executions on each instance. To obtain a fair comparisons with these metaheuristics, we also performed 10 execution of our ILS on each instance. We allowed it to run for just one second, a much faster time than that by \cite{Ribeiro2012} even considering the difference in the speed of the PCs. 

In Table \ref{tabHeuRigAggregate}, aggregate results for groups of 10 instances having the same number of vertices and rigs are reported. For each algorithm, we show the minimum and average solution costs found in the 10 executions performed on each instance. For the ILS, we also show the percentage gap of the best and average ILS solution costs with respect to those found by the three other metaheuristics.
The results clearly indicate that the ILS is very effective: despite the smaller computational time allowed, it finds the same solution costs for the instances with just 50 vertices, and it improves both average and best costs for the larger instances. 

\begin{table}[htb]
\caption{Heuristic solutions comparison for the WRRP on Rig-instances (10 instances per line)}
{\small
\resizebox{\textwidth}{!}{%
\setlength{\tabcolsep}{3pt}
\begin{tabular}{lrrrrrrrrrr}
\toprule
\multirow{2}{*}{{}} & \multicolumn{6}{c}{{\cite{Ribeiro2012} (30 sec. t.lim., 10 runs per inst.)}}                                                   & \multicolumn{4}{c}{{this work (1 sec. t.lim., 10 runs per inst.)}}                                   \\
\cmidrule(lr){2-7} \cmidrule(lr){8-11} 
                               & \multicolumn{2}{c}{{ILS}} & \multicolumn{2}{c}{{CS}} & \multicolumn{2}{c}{{ALNS}} & \multicolumn{4}{c}{{ILS}}                                  \\
                              \cmidrule(lr){2-3} \cmidrule(lr){4-5}
                               \cmidrule(lr){6-7} \cmidrule(lr){8-11}
instance  & {min cost}   & {avg cost}   & {min cost}   & {avg cost}  & {min cost}    & {avg cost}   & {min cost} & {avg cost} & {gapMin} & {gapAvg} \\     %instance  & {cost$_{\mbox{best}}$}   & {cost$_{\mbox{avg}}$}   & {cost$_{\mbox{best}}$}   & {cost$_{\mbox{avg}}$}  & {cost$_{\mbox{best}}$}    & {cost$_{\mbox{avg}}$}   & {cost$_{\mbox{best}}$} & {cost$_{\mbox{avg}}$} & {gap$_{\mbox{best}}$} & {gap$_{\mbox{avg}}$} \\     
\cmidrule(lr){1-1} \cmidrule(lr){2-3} \cmidrule(lr){4-5}
                               \cmidrule(lr){6-7} \cmidrule(lr){8-11}
{50/5}   & 43350.20   & 43496.73   & 43350.20   & 43350.20   & 43350.20   & 43512.71   & 43350.20   & 43350.20   & 0.00  & 0.00  \\
{50/10} & 30454.01   & 30502.64   & 30455.89   & 30454.01   & 30454.01   & 30458.10   & 30454.01   & 30454.80   & 0.00  & 0.00  \\
{100/5} & 124125.02  & 126176.24  & 124149.44  & 125422.79  & 123900.43  & 125601.43  & 123821.30  & 123845.79  & $-$0.06 & $-$1.26 \\
{100/10} & 75637.67   & 76306.00   & 75468.43   & 75956.03   & 75375.72   & 75597.20   & 75373.40   & 75373.40   & 0.00  & $-$0.30 \\
{500/5} & 1775810.82 & 1799316.87 & 1743878.32 & 1760905.11 & 1721219.09 & 1741979.27 & 1701320.95 & 1716807.32 & $-$1.16 & $-$1.45 \\
{50/10} & 924889.93  & 936241.50  & 921988.37  & 928823.13  & 904862.69  & 918117.26  & 887859.35  & 890292.83  & $-$1.88 & $-$3.03\\
                               \cmidrule(lr){1-1} \cmidrule(lr){2-3} \cmidrule(lr){4-5}
                               \cmidrule(lr){6-7} \cmidrule(lr){8-11}
{averages}	&{495711.27}	&{502006.66}	&{489881.77}	&{494151.88}	&{483193.69}	&{489210.99	}&{477029.87}	& {480020.72} &	{$-$1.28}&	{$-$1.88} \\
 \bottomrule
\end{tabular}%
}
}
\label{tabHeuRigAggregate}
\end{table}

\subsection{Results on the wkTRP and Sensitivity Analysis}\label{subsec:wkTRP}
In this section, we present the results that we obtained for the wkTRP and also perform some sensitivity analysis. Table \ref{tabwLQLAgg} reports the results on the wLQL instances obtained by ILS and B\&Cut. The columns have the same meaning of those in Table \ref{tabRig50}, with the exception of gap that now is not reported because all instances are solved to the optimality. Each line gives average values over 10 instances. The last line reports overall averages. It can be noticed that the ILS is, once more, capable of finding the optimal solution value for all instances. This is achieved in less than a second on average. The B\&Cut closes to proven optimality all instances in less than 40 seconds on average and in about 120 seconds in the worst group (brd14051 with $n=40$). The number of explored nodes never exceeds 500. The separation time is very small when compared to the entire execution time, less than a hundredth. Among the cuts, the pigeonhole ones are the largest in number, as more than 3000 violations have been found in practice. The number of $z$- and $f$-activation cuts that have been separated during the enumeration is around 660 and 500, respectively. 
\begin{table}
\caption{Computational results on the wLQL set (10 wkTRP instances per line)}\label{tabwLQLAgg}
\centering
{\small
%\resizebox{\textwidth}{!}{%
% \scriptsize
% \setlength{\tabcolsep}{0.1cm}
\begin{tabular}{lrrrrrrrrrr}
\toprule
\multicolumn{2}{c}{{instance}} & \multicolumn{2}{c}{{ILS}} & \multicolumn{7}{c}{{B\&Cut}} \\
\cmidrule(lr){1-2} \cmidrule(lr){3-4} \cmidrule(lr){5-11} 
{name}     & {$n$} & {cost}      & {time}  & {cost}   & {nodes}    & {time}  & {sep.time} & {\#pgn}     & {\#$z$Act}   & {\#$f$Act}   \\
\cmidrule(lr){1-2} \cmidrule(lr){3-4} \cmidrule(lr){5-11} 
brd14051 & 30 & 108415.33 & 0.28 & 108415.33 & 44.00  & 13.03  & 0.05 & 1195.60 & 402.30  & 320.30 \\
d15112   & 30 & 284114.64 & 0.27 & 284114.64 & 58.70  & 10.46  & 0.06 & 1312.10 & 417.20  & 331.90 \\
d18512   & 30 & 106996.04 & 0.26 & 106996.04 & 0.50   & 6.02   & 0.03 & 1223.10 & 402.50  & 302.00 \\
fnl4461  & 30 & 61191.07  & 0.27 & 61191.07  & 9.80   & 4.42   & 0.02 & 1106.00 & 340.10  & 249.10 \\
nrw1379  & 30 & 37690.96  & 0.27 & 37690.96  & 124.20 & 20.40  & 0.13 & 1340.00 & 441.20  & 356.40 \\
pr1002   & 30 & 204226.26 & 0.26 & 204226.26 & 33.80  & 8.00   & 0.04 & 1220.80 & 414.30  & 307.40 \\
brd14051 & 40 & 126306.84 & 0.44 & 126306.84 & 367.10 & 120.34 & 0.77 & 3165.90 & 893.60  & 737.50 \\
d15112   & 40 & 358816.21 & 0.45 & 358816.21 & 0.60   & 6.42   & 0.03 & 2437.10 & 542.00  & 399.20 \\
d18512   & 40 & 138725.90 & 0.41 & 138725.90 & 75.40  & 24.31  & 0.18 & 2880.80 & 670.50  & 518.60 \\
fnl4461  & 40 & 76687.52  & 0.46 & 76687.52  & 42.50  & 21.21  & 0.16 & 2606.70 & 592.40  & 438.30 \\
nrw1379  & 40 & 44836.65  & 0.46 & 44836.65  & 109.00 & 23.40  & 0.21 & 2520.70 & 497.80  & 419.20 \\
pr1002   & 40 & 266619.69 & 0.44 & 266619.69 & 39.40  & 25.74  & 0.16 & 2842.60 & 704.60  & 492.40 \\
brd14051 & 50 & 159048.44 & 0.68 & 159048.44 & 42.30  & 30.04  & 0.29 & 4631.20 & 714.80  & 594.60 \\
d15112   & 50 & 446461.43 & 0.70 & 446461.43 & 173.40 & 86.20  & 1.09 & 6305.20 & 1049.00 & 755.00 \\
d18512   & 50 & 167873.36 & 0.67 & 167873.36 & 194.50 & 63.23  & 0.78 & 5455.30 & 1042.70 & 719.70 \\
fnl4461  & 50 & 97877.68  & 0.69 & 97877.68  & 21.60  & 27.72  & 0.20 & 4366.20 & 717.30  & 546.80 \\
nrw1379  & 50 & 57229.84  & 0.66 & 57229.84  & 451.80 & 114.69 & 1.77 & 6224.60 & 1168.20 & 858.80 \\
pr1002   & 50 & 331764.46 & 0.70 & 331764.46 & 41.00  & 60.17  & 0.37 & 5034.20 & 853.50  & 608.60 \\
\cmidrule(lr){1-2} \cmidrule(lr){3-4} \cmidrule(lr){5-11} 
{averages} &    & 170826.80 & 0.46 & 170826.80 & 101.64 & 36.99  & 0.35 & 3103.78 & 659.11  & 497.54\\
\bottomrule
\end{tabular}
%}

}
\end{table}

The results that we obtained can be also used to asses how computationally relevant is the inclusion of weights in the problem. Table \ref{tb:compW} shows the average B\&Cut processing times in seconds for LQL- and wLQL-instances, for groups of instances having 30, 40 and 50 vertices. The results on the LQL-set are derived from Table \ref{tabwLQLAgg}, and those on the LQL set from Table \ref{tabTSPLibAggregate}. The average solution time increased substantially for instances with 30 and 40 vertices, while it slightly decreased for instances with 50 vertices. The last decrease can be imputed to a couple of instances. On average, the B\&Cut required 17.13\% more time to process the wLQL-instances, so we can conclude that the addition of the weight component makes the problem slightly more difficult.
\begin{table}[htb]
    %\small
    \centering
    \small
    \caption{Average B\&Cut times on weighted (wLQL) and unweighted (LQL) problems (60 instances per line)}\label{tb:compW}
        \begin{tabular}{lrrr}
            \toprule
            {$n$} & {unweighted} & {weighted}  & {increase(\%)} \\ 
            \cmidrule(lr){1-1} \cmidrule(lr){2-4}
            30   & 5.67    & 10.39  & 83.24 \\
            40   & 22.05   & 36.90 & 67.34\\
            50   & 67.02   & 63.68 & -4.98 \\
            \cmidrule(lr){1-1} \cmidrule(lr){2-4}
            averages   & 31.58   & 36.99 & 17.13 \\ 
            \bottomrule& 
        \end{tabular}
\end{table}

\begin{figure}[htb]
    \centering
    \includegraphics[width=0.6\linewidth]{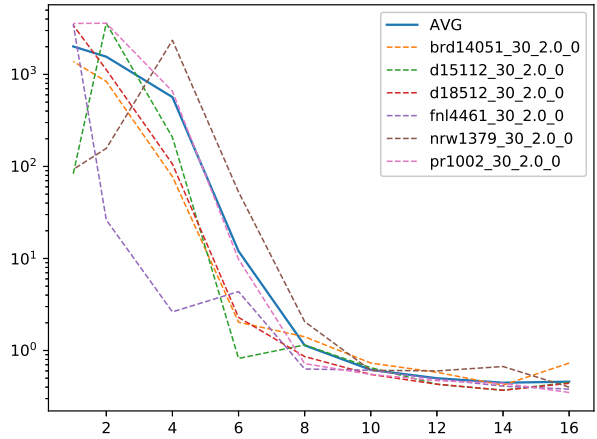}
    \caption{B\&Cut execution times in seconds for $k =  1, 2, 4, 6, 8, 10, 12, 14, 16$} \label{fig:grafico1}
\end{figure}

The next test we performed aimed at assessing the impact of the number $K$ of repairmen. Figure \ref{fig:grafico1} shows the CPU times needed by the B\&Cut to solve the first instance of each group having the same seminal routing instance. We limited the test to instances with 30 vertices, imposed just one hour of time limit, and tested different values for $K = 1, 2, 4, 6, 8, 10, 12, 14, 16$.  The figure shows that the smaller the number of repairmen is, the harder the instance becomes. Indeed, we observed that for $K=2$ two instances were unsolved in the one-hour time limit (d15112 and pr1002). All other instances were solved. The execution times were always below 1 seconds when $K \geq 10$.

To better understand the importance of each proposed inequality, we ran several variations of the B\&Cut algorithm, in each of which we suppressed one cut at a time. 
We also performed tests in which we suppressed the ILS starting solution and the first phase of the B\&Cut.

Table \ref{tabVersions} shows the obtained results by summarizing them for groups of instances having the same value of $n$. The B\&Cut variants have been sorted from the fastest one (i.e., the complete one) to the slowest one. All instances were solved to the proven optimality under all variants, still an important increase in the solution time can be observed when one of the component is removed from the algorithm. The removal of $f$- or $z$-activation cuts increases the average time from 36.99 seconds to 46.38 or 56.82, respectively. Removing the ILS has a stronger effect and increases the average time to 70.63 seconds, almost doubling the effort required by the complete B\&Cut version. Suppressing the first phase has an even stronger effect, especially on the larger instances with 50 vertices, where the time gets almost five times larger. 
The importance of the Pigeonhole cuts in the solution process is evident, as their suppression causes an increase of more than 10 times in the average CPU time. 

\begin{table}[htb]
\caption{Average times for different B\&Cut versions on the wLQL set (60 instances per line)}
\centering
\small
\begin{tabular}{lrrrrrr}
\toprule
 & \multicolumn{6}{c}{{B\&Cut version}} \\ 
 \cmidrule(lr){2-7}
$n$	&	complete 	&	no $f$Act 	&	no $z$Act	&	no ILS  	&	no 1st phase 	&	no pigeon\\               
\cmidrule(lr){1-1} \cmidrule(lr){2-7}
30  & 10.38 & 12.27 & 17.34 & 18.49  & 19.48  & 148.73  \\
40  & 36.90 & 48.96 & 62.31 & 68.29  & 208.64 & 604.68  \\
50  & 63.68 & 77.91 & 90.82 & 125.13 & 356.69 & 1126.31 \\
\cmidrule(lr){1-1} \cmidrule(lr){2-7}
averages	&	 36.99 & 46.38 & 56.82 & 70.63  & 194.94 & 626.57 \\
\bottomrule
\end{tabular}
%}
\label{tabVersions}
\end{table}

\section{Conclusions}\label{sec:conclusions}
In this work, we proposed a generalization of the well-known traveling repairman problem, called the weighted $k$-traveling repairman problem (wkTRP), that calls for  routing  a  set  of  technicians to visit customers by minimizing the total weighted latency, that is, the sum of the weighted waiting time of each customer. 
The wkTRP has not been previously addressed in the literature, but we consider it a problem of interest to researchers and practitioners because it is a quite general problem that models many real-life scenarios. In particular, it allowed us to model the process of scheduling the maintenance of speed cameras in the metropolitan area of Rio de Janeiro, Brazil.

We proposed a mathematical formulation and enriched it with a number of families of cutting planes. We consequently developed a tailored branch-and-cut (B\&Cut) algorithm, in addition to an Iterated Local Search (ILS) heuristic to provide the B\&Cut with initial solutions. Incidentally, the ILS became a relevant contribution of this work, as it achieved excellent practical results, with high-quality solutions and low computing times.

The two algorithms performed very well on the Rio de Janeiro instances, consistently improving the manual solutions produced by the company and demonstrating the practical importance of this study. We remark that both algorithms can easily be adapted to cope with other problems from the literature, raising the question of whether they also compare favorably with existing optimization methods. We believe our extensive computational results have provided a positive answer to this.

There are several interesting future research directions that we believe should be pursued. Since the wkTRP (and, in general, most problems related to the TRP) is very difficult, it would be relevant to work on the development of more advanced exact solution methods, especially for the solution of large-scale instances. Based on the large number of results from the literature, B\&Cut algorithms appear to the most promising technique.
As the results on our case study data demonstrated, there can be important differences between travel and operating times that are expected from average historical values and those that are incurred in the daily operations. These variations can be taken into account by specific stochastic algorithms, which, however, tend to converge very slowly to optimal solutions. This is also another interesting research direction.

In traveling repairman problems, it is always assumed that each customer is visited once. It would be interesting, instead, to study cases in which multiple visits are allowed. This can be important to model, for example, inventory routing problems (\cite{ABLS07}) and pickup-and-delivery problems with split deliveries  (\cite{BI17}), such as those arising in bike sharing rebalancing problems (\cite{BCIS19}). In this case, some bike stations are more used than others and thus delays at servicing them have a large impact, i.e., a larger weight on the latency should be considered. In a dynamic scenario, rebalancing might be required multiple times a day, hence requiring multiple visits.
Finally, the inclusion of weights in the problem description made the problem more general, allowing a wider range of real-world applications to be modeled. This was obtained at the expense of a limited increase in the computational effort required by the solution algorithms. Thus, it appears  interesting to study other problem variants that include weighted objective functions, such as, e.g., weighted tardiness in case of due dates.
 
%\acknowledgment{%
%    The last author acknowledges financial support from University of Modena and Reggio Emilia under grant FAR 2018 ``Analysis and optimization of healthcare and pharmaceutical logistic processes''.
%}% Leave this (end of acknowledgment)

%\section*{References}
%\small{
%\bibliography{kWTRP}
%}  
\small{
\bibliographystyle{unsrt} % outcomment this and next line in Case 1
\bibliography{kWTRP} % if more than one, comma separated    
}
\end{document}